\documentclass[12pt]{article}%
\usepackage[nosort]{cite}
\usepackage[usenames, dvipsnames]{xcolor}
\usepackage{graphicx}
\usepackage{multicol}
\usepackage{amsfonts}
\usepackage{amssymb}
\usepackage{amsmath}
\usepackage{heck}
\usepackage{afterpage}
\usepackage{setspace}
\usepackage{verbatim}
\usepackage{color}
\usepackage{longtable}
\usepackage{float}
\usepackage{subcaption}
\usepackage{epsfig}
\usepackage{enumerate}
\usepackage{epstopdf}
\usepackage{adjustbox}
\usepackage{multirow}
\usepackage{tikz}
\usepackage[margin=1in]{geometry}
\usepackage{titletoc}
\usepackage{hyperref}
\usepackage[percent]{overpic}
\usepackage{gensymb}
\usepackage{mathtools}
\usepackage{tikz-cd}%
\providecommand{\U}[1]{\protect\rule{.1in}{.1in}}
\pdfoutput=1
\newsavebox{\mysavebox}

\hypersetup{colorlinks,citecolor=black,filecolor=black,linkcolor=black,urlcolor=black}
\usetikzlibrary{decorations.markings}

\numberwithin{equation}{section}

\usetikzlibrary{chains}
\allowdisplaybreaks
\tikzset{node distance=2em, ch/.style={circle,draw,on chain,inner sep=2pt},chj/.style={ch,join},every path/.style={shorten >=4pt,shorten <=4pt},line width=1pt,baseline=-1ex}

\newcommand{\ba}{\begin{eqnarray}}
\newcommand{\ea}{\end{eqnarray}}

\newcommand{\be}{\begin{equation}}
\newcommand{\ee}{\end{equation}}

\tikzstyle{startstop} = [rectangle, rounded corners, minimum width=3cm, minimum height=1cm,text centered, draw=black, fill=blue!10]
\tikzstyle{startstop} = [rectangle, rounded corners, minimum width=3cm, minimum height=1cm,text centered, draw=black, fill=blue!10]
\tikzstyle{io} = [trapezium, trapezium left angle=70, trapezium right angle=110, minimum width=3cm, minimum height=1cm, text centered, draw=black, fill=blue!30]
\tikzstyle{process} = [rectangle, minimum width=3cm, minimum height=1cm, text centered, draw=black, fill=orange!30]
\tikzstyle{decision} = [diamond, minimum width=3cm, minimum height=1cm, text centered, draw=black, fill=green!30]
\tikzstyle{arrow} = [thick,->,>=stealth]
\tikzset{->-/.style={decoration={
  markings,
  mark=at position #1 with {\arrow[scale=2.4]{>}}},postaction={decorate}}}
\makeatletter \@addtoreset{equation}{section} \makeatother

\begin{document}


\date{August 2021}

\title{Misanthropic Entropy and\\[4mm]Renormalization as a Communication Channel}

\institution{VCS}{\centerline{$^{1}$Vigilant Cyber Systems, Mt Airy, NC 27030, USA}}

\institution{PENN}{\centerline{$^{2}$Department of Physics and Astronomy, University of Pennsylvania, Philadelphia, PA 19104, USA}}

\authors{Ram\'{o}n Fowler\worksat{\VCS}\footnote{e-mail: {\tt ramonfowler55@gmail.com}} and
Jonathan J. Heckman\worksat{\PENN}\footnote{e-mail: {\tt jheckman@sas.upenn.edu}}}

\abstract{A central physical question is the extent
to which infrared (IR) observations are sufficient to reconstruct a candidate ultraviolet (UV) completion.
We recast this question as a problem of communication, with messages encoded in field configurations
of the UV being transmitted to the IR degrees of freedom via a noisy channel
specified by renormalization group (RG) flow, with noise generated by coarse graining / decimation.
We present an explicit formulation of these considerations in terms of lattice field theory,
where we show that the ``misanthropic entropy''---the mutual information obtained from decimating neighbors---encodes the extent to which information is lost in marginalizing over / tracing out UV degrees of freedom. Our considerations apply both to statistical field theories as well as density matrix renormalization of quantum systems, where in the quantum case, the statistical
field theory analysis amounts to a leading order approximation. As a concrete example, we focus on the case of the 2D Ising model,
where we show that the misanthropic entropy detects the onset of the phase transition at the Ising model critical point.}

\maketitle

\setcounter{tocdepth}{2}



\newpage

\section{Introduction \label{sec:INTRO}}

The notion of renormalization group flow
from the ultraviolet (UV) to the infrared (IR) is central
to the study of many physical phenomena.
Renormalization provides a deep organizational
principle in which physical phenomena are arranged according
to different length / energy scales \cite{Gell-Mann:1954yli, Kadanoff:1966wm, Wilson:1971bg, Wilson:1971dh}.

An especially important case concerns the study of string theory and its possible low energy manifestations.
On general grounds, one might ask to what extent a set of IR observables can serve
to constrain the content of some candidate UV completion (see e.g. \cite{Heckman:2013kza,
Balasubramanian:2014bfa, Heckman:2016wte, Heckman:2016jud, Fowler:2020rkl, Balasubramanian:2020lux}
as well as \cite{Balasubramanian:1996bn, 2014arXiv1410.3831M, Erdmenger:2020vmo, Halverson:2020trp,
Stout:2021ubb, Erdmenger:2021sot, Erbin:2021kqf}).
This issue is also of general interest in trying to understand the extent
to which information is lost in passing from short to long distance scales.

\begin{figure}[t!]
\centering
\includegraphics[trim={0 7cm 0 7cm},scale=0.5]{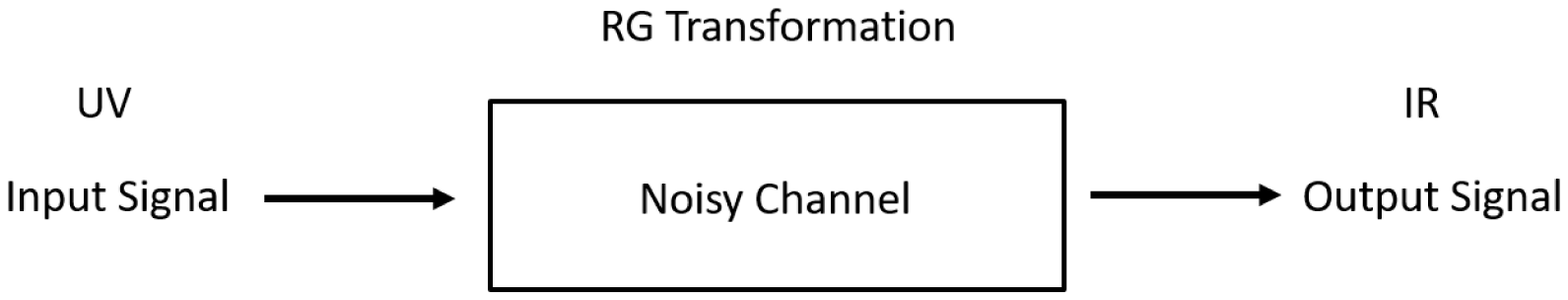}
\caption{Depiction of renormalization as a noisy communication channel.
UV degrees of freedom serve as an input, with IR degrees of freedom as an output.}
\label{fig:UVRGIRChannel}
\end{figure}

Our first aim in this note will be to recast renormalization group
flow as a communication problem in the sense of Shannon \cite{Shannon:1948zz},
where messages encoded in the degrees of freedom of the UV are transmitted via RG flow to
messages encoded in the degrees of freedom of the IR (see Figure \ref{fig:UVRGIRChannel}).
Indeed, for a classical communication channel, one speaks of transmission of messages across a channel $X \rightarrow Y$
involving input random variables $X$ and output random variables $Y$, and this can also be generalized to the quantum setting.
For earlier work discussing UV/IR entanglement, see e.g. \cite{Balasubramanian:2014bfa, Agon:2014uxa}.

Our second aim in this note will be to develop a concrete algorithmic procedure for tracking
this loss of information in terms of local degrees of freedom. To this end, we consider explicit lattice based field theories in which the process of renormalization can be visualized as a grouping of UV variables into a single IR variable, much as in the case of Kadanoff's block spin
renormalization / decimation procedure \cite{Kadanoff:1966wm, Wilson:1971bg, Wilson:1971dh}. Though our considerations apply to a broader class of situations, for ease of exposition we primarily focus on the conceptually simple special case where the degrees of freedom are arranged on a bipartite lattice, so that they can be labelled as black or white. Here, a very natural question is to quantify the mutual information between the black and white degrees of freedom. Since it involves an entropic quantity obtained from decimation of one's neighbors, we refer to this as the ``misanthropic entropy'' $\mathcal{S}_{\mathrm{mis}}$. This notion is quite broad and can also be applied to $n$-partite systems.

The misanthropic entropy is closely related to other notions of proximity one might use in tracking the loss of information during an RG flow. For example, in a statistical field theory, the Boltzmann distribution defines a probability distribution for a given field configuration via $p[\phi] \sim \exp(-S[\phi])$, with $S[\phi]$ the action evaluated on the field configuration $\phi$. Instead of directly thinning out the field theory degrees of freedom, we can instead visualize an RG flow as a trajectory in the space of couplings, and so we can think of a sequence of actions $S_{k}[\phi]$, such that $k = 0$ is in the deep UV, and $k \rightarrow \infty$ indicates the passage to the deep IR. Along this flow, we can track the relative entropy (i.e. the Kullback-Leibler divergence \cite{Kullback:1951}) before and after decimation:
\begin{equation}
\mathcal{S}_{\mathrm{flow}}^{(k - 1 || k)} = \int d\phi \, p_{k - 1} \mathrm{\log} \frac{p_{k - 1}}{p_{k}}.
\end{equation}
In the limit where each decimation step removes an infinitesimal number 
of degrees of freedom (i.e. integrating only a thin momentum shell), we have:
\begin{equation}
\mathcal{S}^{(k - 1||k )}_{\mathrm{mis}}(\mathrm{inf}) = \mathcal{S}_{\mathrm{flow}}^{(k - 1||k )}(\mathrm{inf}) + \mathcal{O}(\delta \lambda^2),
\end{equation}
where $\delta \lambda$ is the infinitesimal change in continuum field theory couplings upon performing the decimation step.
Said differently, in the infinitesimal limit, different notions of renormalization and communication agree. Integrating these infinitesimal quantities can produce different notions of comparison between UV degrees of freedom and their decimated counterparts,
which we label similarly as $\mathcal{S}^{(\mathrm{UV} || k)}$, and this is a priori different from a sum over the individual $\mathcal{S}^{(k-1 || k)}$. For other notions of distinguishability, especially in the context of pattern recognition in complex systems, see, e.g., \cite{Bagrov30241, sotnikov2021certification, PhysRevX.10.011037, gokmen2021symmetries}.

To illustrate how this works in practice, we also present an explicit example based on the 2D Ising model statistical field theory.
In particular, we show that as we continue to decimate the original lattice, the approach to the critical value of the coupling is detected by an abrupt increase in the misanthropic entropy $\mathcal{S}_{\mathrm{mis}}^{(\mathrm{UV}|| k)}$ due to the increased mutual information stored in long range correlations.

In fact, we can also use $\mathcal{S}^{(\mathrm{UV}||k)}_{\mathrm{mis}}$ as a diagnostic to detect the possible presence of additional operators
in an effective field theory generated by decimation. Indeed, in the 2D Ising model, we find that if one neglects the next to nearest neighbor interactions generated by decimation, then a naive evaluation of the misanthropic entropy obtained by dropping higher order interactions
would produce a negative number, something which is impossible for the relative entropy. Said differently, demanding non-negativity of  $\mathcal{S}^{(\mathrm{UV}||k)}_{\mathrm{mis}}$ \textit{predicts} the appearance of specific structure in the decimated theory.

For the flow entropy $\mathcal{S}_{\mathrm{flow}}^{(\mathrm{UV}|| k)}$, the value at the critical
point vanishes. Small perturbations away from this point
also produce more pronounced divergences at large decimation step because the 2D
Ising model has an unstable fixed point.

\textbf{Note:} The present work is based in part on material which appeared in the UNC Chapel Hill PhD thesis by one of us \cite{Fowler:2020rkl}.
As we were preparing this note for submission to the pre-print \texttt{arXiv}, reference \cite{Erdmenger:2021sot} appeared
which studies related but distinct notions of information loss under RG flow.

\section{Renormalization as a Communication Channel}

We now show that renormalization of a statistical field theory can be viewed as specifying a communication channel.
To frame the discussion to follow, recall that for a classical communication channel $X \rightarrow Y$, we view $X$ and
$Y$ as random variables (for additional review, see e.g. \cite{Cover2006}). There is a corresponding joint probability distribution:
\begin{equation}
p_{X,Y}(x,y) = p_{Y|X}(y|x) p_{X}(x),
\end{equation}
which captures the sense in which the random variables $X$ and $Y$ inform one another. For example,
the mutual information is obtained from the relative entropy:
\begin{equation}
I(X,Y) = \int dx dy \, p_{X,Y} \log \frac{p_{X,Y}}{p_X p_Y},
\end{equation}
where the distribution $p_{Y}(y)$ is obtained by marginalizing $p_{X,Y}(x,y)$ over $x$.
Maximizing over candidate $p_X$ (subject to suitable constraints) defines the
channel capacity, and minimizing over candidate $p_{Y|X}$ subject to some rate of fidelity determines
the optimal distortion rate for a channel. There are also quantum analogs of these notions, where we replace
the probability distributions by suitably prepared density matrices and measurement operations (see e.g. \cite{2012RvMP...84..621W}
for a review).

Now, in the context of renormalization, it is tempting to view $X$ as the UV degrees of freedom, and $Y$ as the IR degrees of freedom
of a field theory. In the case of a statistical field theory, we can (much as in \cite{Balasubramanian:2014bfa}) specify a probability
distribution for field configurations using the Boltzmann factor:
\begin{equation}
p[\phi] = \frac{1}{Z} \exp(-S[\phi]),
\end{equation}
where $S[\phi]$ is the action, and $Z$ is the partition function.

The content of a renormalization group transformation can be specified in terms of a conditional distribution $p_{\mathrm{IR|UV}}[\phi_{\mathrm{IR}} | \phi_{\mathrm{UV}}]$
which amounts to regrouping the UV degrees of freedom in terms of IR degrees of freedom:
\begin{equation}
p_{\mathrm{IR}}[\phi_{\mathrm{IR}}] = \int d\phi_{\mathrm{UV}} \, p_{\mathrm{IR|UV}}[\phi_{\mathrm{IR}} | \phi_{\mathrm{UV}}] p_{\mathrm{UV}}[\phi_{\mathrm{UV}}],
\end{equation}
where the integral over $\phi_{\mathrm{UV}}$ is really a path integral of all possible field configurations.
To a continuum field theorist, one might prefer to just set $p_{\mathrm{IR|UV}} = 1$, and view this operation as ``integrating out momentum shells.'' On the other hand, the present formulation allows us to explicitly define a notion of block spin renormalization in the sense of Kadanoff, and also applies to setups where the UV degrees of freedom might be packaged rather differently from how they appear in the IR.\footnote{For example, the high and low energy limits of QCD.}
In any case, we can then define a joint distribution:
\begin{equation}
p_{\mathrm{UV,IR}}[\phi_{\mathrm{UV}},\phi_{\mathrm{IR}}] = p_{\mathrm{IR|UV}}[\phi_{\mathrm{IR}}|\phi_{\mathrm{UV}}] p_{\mathrm{UV}}[\phi_{\mathrm{UV}}],
\end{equation}
and the decimation procedure defined by $p_{\mathrm{IR|UV}}$ specifies
a communication channel, which transmits information from the UV to the IR.

The mutual information between UV and IR degrees of freedom is given by:\footnote{Let us emphasize here that even though $p_{\mathrm{UV}}$ can be obtained by integrating $p_{\mathrm{UV,IR}}$ over the IR variables, this does not produce a non-local effective action.}
\begin{equation}
I(\mathrm{UV,IR}) = \int d\phi_{\mathrm{UV}} d\phi_{\mathrm{IR}} \, p_{\mathrm{UV,IR}} \mathrm{log} \frac{p_{\mathrm{UV,IR}}}{p_{\mathrm{UV}}p_{\mathrm{IR}}}.
\end{equation}
A next natural question might be to study the channel capacity of this system,
as given by maximizing $I(\mathrm{UV,IR})$ over candidate $p_{\mathrm{UV}}$'s subject to suitable constraints.
One can also study the distortion rate by instead minimizing over possible channels (i.e. block spin transformations)
$p_{\mathrm{IR|UV}}$.

It is also important to understand the extent to which the UV and IR theories are truly distinguishable.
Using the formalism of \cite{Balasubramanian:2014bfa}, we expect this to be encoded in a relative entropy between the probability distributions for the UV and IR variables. The present formulation in terms of a communication channel does not immediately afford us with a way to study this question, because strictly speaking, $\phi_{\mathrm{UV}}$ and $\phi_{\mathrm{IR}}$ are just different bases of fields. Rather, one could instead speak of two different UV distributions $p_{\mathrm{UV}}$ and $p^{\prime}_{\mathrm{UV}}$, and then compare the mutual information stored in the channel in these two situations.

Indeed, in the context of a continuum field theory, we can visualize an RG flow as a trajectory in the space of couplings \cite{Polchinski:1983gv}. Then, we can keep the same basis of fields, and for each one speak of a probability distribution for field configurations $p[\phi] \sim \exp(-S[\phi])$, as well as the corresponding relative entropy between these distributions.
Breaking up the flow into a set of discretized steps gives us a sequence of actions $S_{k}[\phi]$ so
we can consider a corresponding ``flow entropy'':
\begin{equation}
\mathcal{S}_{\mathrm{flow}}^{(i||j)} \equiv \int d \phi \, p_{i} \mathrm{log} \frac{p_i}{p_{j}}.
\end{equation}
Two special cases of interest are $\mathcal{S}_{\mathrm{flow}}^{(k-1||k)}$ and $\mathcal{S}_{\mathrm{flow}}^{(\mathrm{UV}||k)}$.
The former tells us about a local change in proximity and the latter tells us about
an integrated notion of proximity from the original UV distribution. It is also convenient
to introduce a measure of entropy density:
\begin{equation}
s_{\mathrm{flow}} \equiv \mathcal{S}_{\mathrm{flow}} / N,
\end{equation}
where $N$ is the total number of lattice sites.

The relation between the communication channel formulation and the flow entropy is not immediately obvious. For example, the continuum field theory answer involves the \textit{same} field theory degrees of freedom. This is traditionally viewed as integrating out momentum shells, and then rescaling the spacetime so as to have the same domain of support for the physical fields. This last step is somewhat more subtle to define as an operation on a communication channel, but can be viewed as essentially re-encoding the original message in terms of the original set of variables, much as in \cite{Tishby:2000bn}.

To further study this issue, we now turn to a quantity which makes contact with both notions.

\section{Misanthropic Entropy}

We now introduce a proxy for tracking the mutual information of local degrees of freedom under renormalization group flow.
In general terms, we can view the process of renormalization as splitting up our ultraviolet degrees of freedom into (at least) two
subsets, and then integrating out (in statistical field theory) or tracing out (in quantum statistical ensembles) such that
the remaining degrees of freedom retain a suitable notion of locality. For further details on the more abstract treatment of these statements in the context of operator algebras of lattice quantum field theories, see e.g. \cite{Radicevic:2016kpf} as well as
\cite{Radicevic:2021ykf, Radicevic:2021zcz, Radicevic:2021wty}.

To avoid unnecessary complications, our main focus here will be on the case of statistical field theories where the degrees of freedom
are localized on a bipartite lattice such that after decimation, we can continue to partition up the degrees of freedom in the same way (i.e. the decimated lattice is also bipartite). Since we are dealing with a bipartite lattice,
we can partition the UV degrees of freedom $\phi_{\mathrm{UV}}$ into those which are localized on
black sites, and those which are localized on white sites, namely $\phi_b$ and $\phi_w$, respectively.
The process of decimation amounts to evaluating:
\begin{equation}
p_{\mathrm{IR}}[\phi_{\mathrm{IR}}] = \int d\phi_b d\phi_w \, \delta(\phi_{\mathrm{IR}} - \phi_{b}) p_{b,w}[\phi_{b}, \phi_w],
\end{equation}
which is just a specific choice of communication channel:
\begin{equation}
p_{\mathrm{IR}|\mathrm{UV}}(\phi_{\mathrm{IR}} | \phi_{\mathrm{UV} = \{b,w\}}) = \delta(\phi_{\mathrm{IR}} - \phi_{b}).
\end{equation}
Observe that for this choice of channel, the mutual information $I(\mathrm{UV},\mathrm{IR})$ is just:
\begin{equation}
I(\mathrm{UV}, \mathrm{IR}) = \mathcal{S}_{b},
\end{equation}
i.e. the entropy of the black site theory. In an actual lattice field theory calculation, one might consider softening the delta function, or by taking various ``majority rules'' for the choice of the averaged block spin, but again we primarily stick to the simplest choice to illustrate the main points.

By abuse of notation, we shall usually just write $p_b[\phi_b]$ to denote the distribution obtained by marginalizing over the complementary degrees of freedom. We could equally well have considered marginalizing over the black sites instead, and so we can also introduce $p_{w}[\phi_{w}]$. The mutual information between the black and white sites defines the misanthropic entropy:
\begin{equation}
\mathcal{S}_{\mathrm{mis}} \equiv I(b,w) = \int d\phi_b d\phi_w \, p_{b,w} \mathrm{log} \frac{p_{b,w}}{p_b p_w},
\end{equation}
which we can also express in terms of the Shannon entropies:
\begin{equation}\label{eqn:smisdef}
\mathcal{S}_{\mathrm{mis}} = - \mathcal{S}_{b,w} + \mathcal{S}_{b} + \mathcal{S}_{w} = - \mathcal{S}_{b,w} + 2\mathcal{S}_{b},
\end{equation}
where in the last equality, we used the assumption that there exists a black / white symmetry.

Summarizing, we are computing the mutual information between nearest neighbors after a single step of decimation. But now we can consider repeating this procedure to obtain for each $k$ the entropies $\mathcal{S}^{(k-1||k)}_{\mathrm{mis}}$, which compares the ($k-1$)-decimated theory to its $k$-decimated counterpart. We can also consider the entropies $\mathcal{S}^{(\mathrm{UV}||k)}_{\mathrm{mis}}$ which involves taking the relative entropy of the original UV distribution (at $k = 0$) with the product distribution obtained from $2^{k}$
copies of the decimated theory.

Much as for the flow entropy, it is convenient to consider the
entropy density:
\begin{equation}
s_{\mathrm{mis}} \equiv \mathcal{S}_{\mathrm{mis}} / N.
\end{equation}
Note that although after $k$ decimation steps we have $N(k) = N / 2^k$ remaining lattice sites, the misanthropic entropy
still makes reference to all $N$ lattice sites.

There are various generalizations we can entertain. For example, it is important to consider how local operators change under RG flow. To do this, we could add source terms for local operator insertions in a UV Hamiltonian, and then generate a sequence of Hamiltonians after decimation. We could then check how the entropy changes after a decimation/RG step. We could also have chosen to perform a finer partitioning of our degrees of freedom. Suppose, for example, that we have an $n$-partite lattice. Then, for a distribution which depends on these colors $p_{1,...,n}$, we can marginalize over all but one and compute the relative entropy:
\begin{equation}
\mathcal{S}_{\mathrm{mis,gen}} = \int d\phi_1 ... d\phi_n \, p_{1,...,n} \mathrm{log} \frac{p_{1,...,n}}{p_1 ... p_n},
\end{equation}
which provides a related notion of misanthropy.

A priori, there is also no need to partition up the system to equal numbers of black and white sites. For example, in an $n$-partite lattice with $n$ very large, we might instead form a new subsystem comprised of the distributions $p_{1}$ and $p_{2,...,n}$, in the obvious notation. In this case, marginalizing over color $1$ leads to a very small modification of the original distribution. More precisely the action for the original
theory and the one obtained from decimating one color are related as:
\begin{equation}
S_{1,...,n} = S_{2,...,n} + S_{1} + S_{\mathrm{mix}},
\end{equation}
where $S_{1}$ and $S_{\mathrm{mix}}$ are both viewed as small perturbations.
This is just the real space version of ``integrating out a momentum shell''.
On the other hand, after performing this decimation step, we can consider
a closely related action as obtained by just working in terms of the original basis of fields:
\begin{equation}
S_{1,...,n} = S_{1,...,n}^{\prime} + \delta S,
\end{equation}
namely we just consider a motion in the space of couplings. \footnote{These fields are actually the renormalized fields, not the bare fields. 
The sense in which we are considering a motion in the space of couplings here is that of Polchinski's exact RGE \cite{Polchinski:1983gv}.} Said differently, we have:
\begin{equation}
S_{1,...,n}^{\prime} = S_{2,...,n} + S_{1} + \mathcal{O}(\delta \lambda),
\end{equation}
where $\delta \lambda$ is again the change in couplings upon performing the decimation step and the $\mathcal{O}(\delta \lambda)$ terms are subleading corrections. In this limit, we observe that the misanthropic and flow entropies agree:
\begin{equation}
\mathcal{S}^{(k - 1 || k )}_{\mathrm{flow}}(\mathrm{inf}) = \mathcal{S}^{(k - 1 || k )}_{\mathrm{mis}}(\mathrm{inf}) + \mathcal{O}(\delta \lambda^2),
\end{equation}
where the argument ``inf'' serves to remind us that this is really an infinitesimal version of our real space decimation procedure.

\section{Quantum Generalizations}

Though our primary focus will be on classical statistical field theories, we emphasize that the
relative entropy considered above is also the leading order
effect in various quantum generalizations. To see why, suppose we are given a pair of density matrices $\rho \sim \exp(-\widehat{H})$ and $\rho^{\prime} \sim \exp(-\widehat{H}^{\prime})$ such that their commutator has small operator norm:
\begin{equation}
[\rho,\rho^{\prime}] = \delta \,\,\, \text{with} \,\,\, \underset{\overrightarrow{v}}{\mathrm{sup}} \frac{\vert \vert \delta \cdot \overrightarrow{v} \vert \vert}{\vert \vert \overrightarrow{v}\vert \vert} \ll 1.
\end{equation}
While there can sometimes be obstructions to a pair of nearly commuting matrices having close norms, for density matrices, this subtlety is less of a worry \cite{Hastings_2009, glebsky2010commuting}. Consequently, the leading order contribution to the relative entropy is captured by:
\begin{equation}
\mathcal{S}(\rho || \rho^{\prime}) = \mathrm{Tr} \rho \log \rho - \rho \log \rho^{\prime} \simeq D_{KL} (p||q) + \mathcal{O}(\delta),
\end{equation}
where $D_{\mathrm{KL}} (p||q)$ is the classical relative entropy (i.e. the Kullback-Leibler divergence \cite{Kullback:1951}) as obtained by working in the approximation where $\widehat{H}$ and $\widehat{H}^{\prime}$ are simultaneously diagonalizable, in which case we can introduce a corresponding probability distribution $p(E) \sim \exp(-E)$, with $\{E \}$ the eigenvalues of $\widehat{H}$.

Specific quantities such as the misanthropic entropy also generalize to the quantum setting. Consider a quantum theory with local degrees of freedom arranged on a bipartite lattice of the same sort considered above. In this case, we assume that the full Hilbert space is a tensor product of the form:
\begin{equation}
\mathcal{H}_{b,w} \simeq \mathcal{H}_b \otimes \mathcal{H}_w.
\end{equation}
Given a density matrix $\rho_{b,w}$ on the full Hilbert space, we can consider performing a partial trace via decimation:
\begin{equation}
\rho_{b} \equiv \mathrm{Tr}_{\mathcal{H}_w} \rho_{b,w}.
\end{equation}
We can then form a new density matrix on the UV Hilbert space $\mathcal{H}_{b,w}$ by taking $\rho_{b} \otimes \rho_w$.
The quantum misanthropic entropy is then defined as:
\begin{equation}
\mathcal{S}_{\mathrm{mis}} \equiv \mathrm{Tr}_{\mathcal{H}_{b,w}} \rho_{b,w} \log \rho_{b,w} - \mathrm{Tr}_{\mathcal{H}_{b}} \rho_{b} \log \rho_{b} - \mathrm{Tr}_{\mathcal{H}_{w}} \rho_{w} \log \rho_{w}.
\end{equation}

Following a similar set of steps to what one would consider in
the density matrix renormalization group \cite{White:1992zz} and multiscale entanglement renormalization algorithm (MERA) \cite{PhysRevLett.101.180503} (see \cite{2009arXiv0912.1651V} for a helpful review),
we can consider starting from $\rho_{b,w} = \exp(-\widehat{H}_{b,w})$ with $\widehat{H}_{b,w}$
a Hamiltonian operator, and then forming a sequence of decimated Hamiltonians $\widehat{H}^{(k)}_{b,w}$.
For the Hamiltonian $\widehat{H}_{b}$ obtained from tracing out the white degrees of freedom, we
can repackage it in terms of the original lattice sites by applying an isometry
map\footnote{Recall that an isometry $W: \mathcal{H}^{\prime} \rightarrow \mathcal{H}$
is a linear map such that $W^{\dag} W = \mathbf{id}_{\mathcal{H}^{\prime}}$ and $W W^{\dag} = P$ is a projection, i.e. $P^{2} = P$.}
$W: \mathcal{H}_{b} \rightarrow \mathcal{H}_{b,w}$ followed by a ``disentangling'' unitary transformation $U$:
\begin{equation}
\widehat{H}_{b} \rightarrow W^{\dag} \widehat{H}_{b} W \rightarrow U^{\dag} W^{\dag} \widehat{H}_{b} W U = \widehat{H}_{b,w}^{\prime}.
\end{equation}
This provides us with a sequence of density matrices $\rho_{b,w}^{(k)} \sim \exp(- \widehat{H}^{(k)}_{b,w})$, and we can compute the relative entropy between these mixed states. Insofar as $\rho^{(k-1)}_{b,w}$ and $\rho^{(k)}_{b,w}$ nearly commute,
we can again use the related statistical field theories to compute the distinguishability of the two theories as we proceed along an RG flow. We can also make reference to the relative entropy between the original $\rho^{(0)}_{b,w}$ and $\rho^{(k)}_{b,w}$.\footnote{Let us note that in DMRG and MERA, one of the goals is to consider a coarse graining procedure such that the evaluation of operator correlation functions with respect to the original UV ground state can be repackaged in terms of related operator correlation functions in the coarse grained system, i.e. one seeks to enforce $\mathrm{Tr}(\rho O_1 ... O_n) = \mathrm{Tr}(\rho^{\prime} O_1^{\prime} ... O_{n}^{\prime})$. Here, our interest is somewhat different.}

Quantities such as the flow and misanthropic entropy can thus be defined and analyzed, both for statistical field theories and quantum statistical systems. For this reason, we primarily focus on the former case, since in many cases it is the leading order contribution to
the quantum case anyway.

\section{Example: 2D Ising Model}

In this section we illustrate the above considerations with the 2D Ising model with spins
$\sigma_i \in \{ - 1 , +1 \}$ arranged on a bipartite square lattice with periodic boundary conditions.
With respect to the Boltzmann action $S = \beta H$, the Hamiltonian of the model is given by:
\begin{equation}
H = - J \sum_{n.n.}\sigma_i\sigma_j.
\end{equation}
\begin{figure}[t!]
\centering
\includegraphics[trim={0 4cm 0 4cm},scale=0.5]{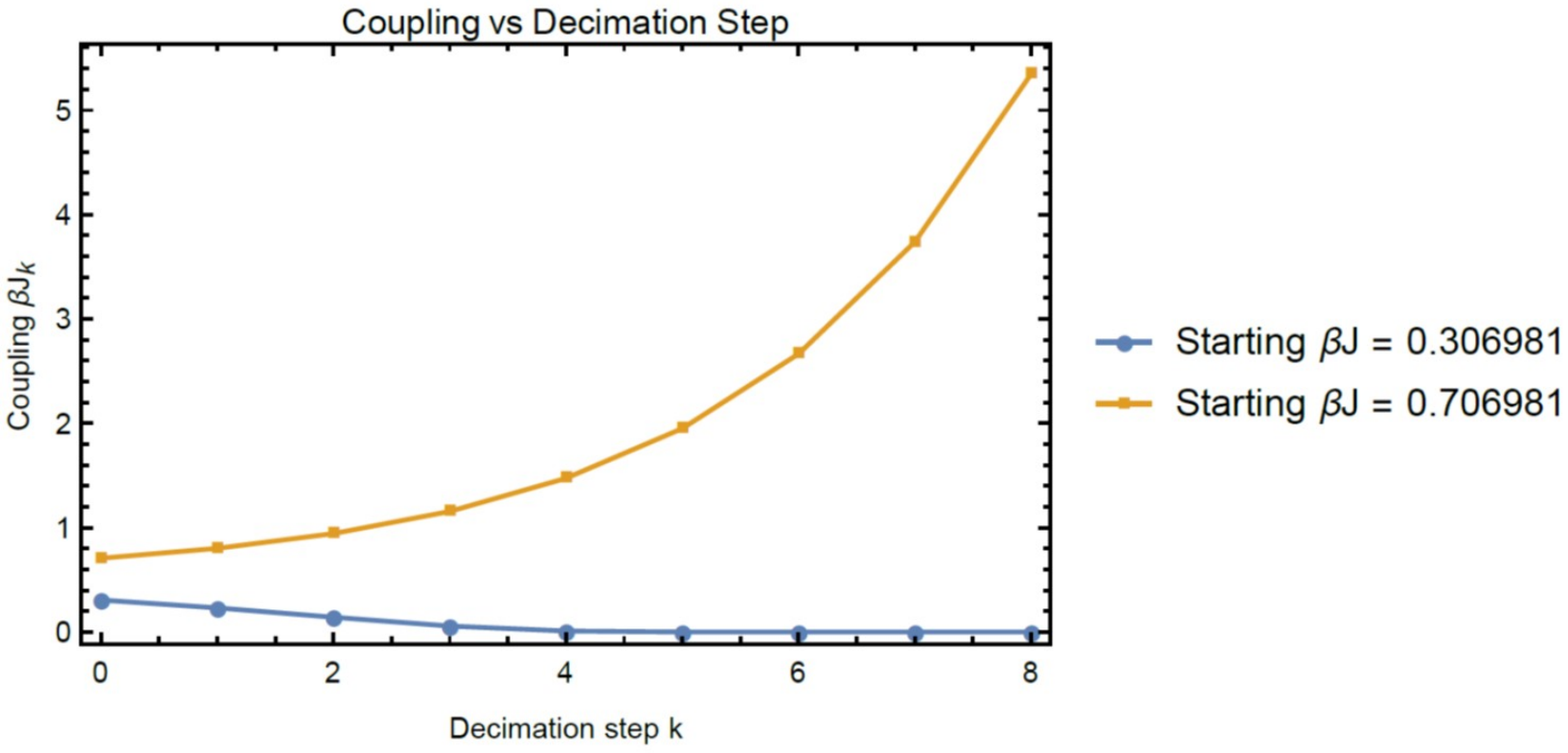}
\caption{The change of the coupling $\beta J_k$ under $k = 8$ decimation steps at two example starting values of $\beta J$: one above the critical point and one below. Above the critical point, $\beta J_k$ grows exponentially. Below the critical point, the $\beta J_k$ decays to zero. The relation for the coupling is obtained assuming that quartic and higher order terms are irrelevant and dropped. The critical point in this approximation scheme is $\beta J_{\ast} \approx 0.50698$.}
\label{fig:couplingvsk}
\end{figure}
After decimation of the white sites, we get a new Hamiltonian, with a shifted value of the nearest neighbor interactions, as well as some additional quartic interaction terms. We label the sequence of nearest neighbor couplings as $J_k$ as obtained from $k$ decimation steps and the corresponding Hamiltonian as $H_k$. In principle, we should also consider the behavior of the other higher order interaction terms, but for the most part we shall omit these contributions. To a certain extent, this approximation is justified because the quartic term is irrelevant at long distances, but the location of $\beta J_{\ast}$ the fixed point of the RG flow (i.e. the critical point)
shifts from the approximate value of $0.51$ to the exact value of $0.44$ (see reference \cite{PhysRev.60.252}). See Figure \ref{fig:couplingvsk} for a plot of $\beta J_k$ versus $k$ at example values below and above the critical value.

Additionally, whereas the entropy $\mathcal{S}_{\mathrm{flow}}^{(\mathrm{UV} || k)}$ is always positive since we are computing a relative entropy between two distributions, neglecting the contribution from the quartic couplings can sometimes produce a negative value for the approximation of the misanthropic entropy, which we write as $\mathcal{S}^{(\mathrm{UV} || k)}_{\mathrm{mis,naive}}$.
This can actually be used as a \textit{diagnostic} to detect the
presence of higher order interaction terms.

\begin{figure}[t!]
\centering
\includegraphics[trim={0 4cm 0 4cm},scale=0.5]{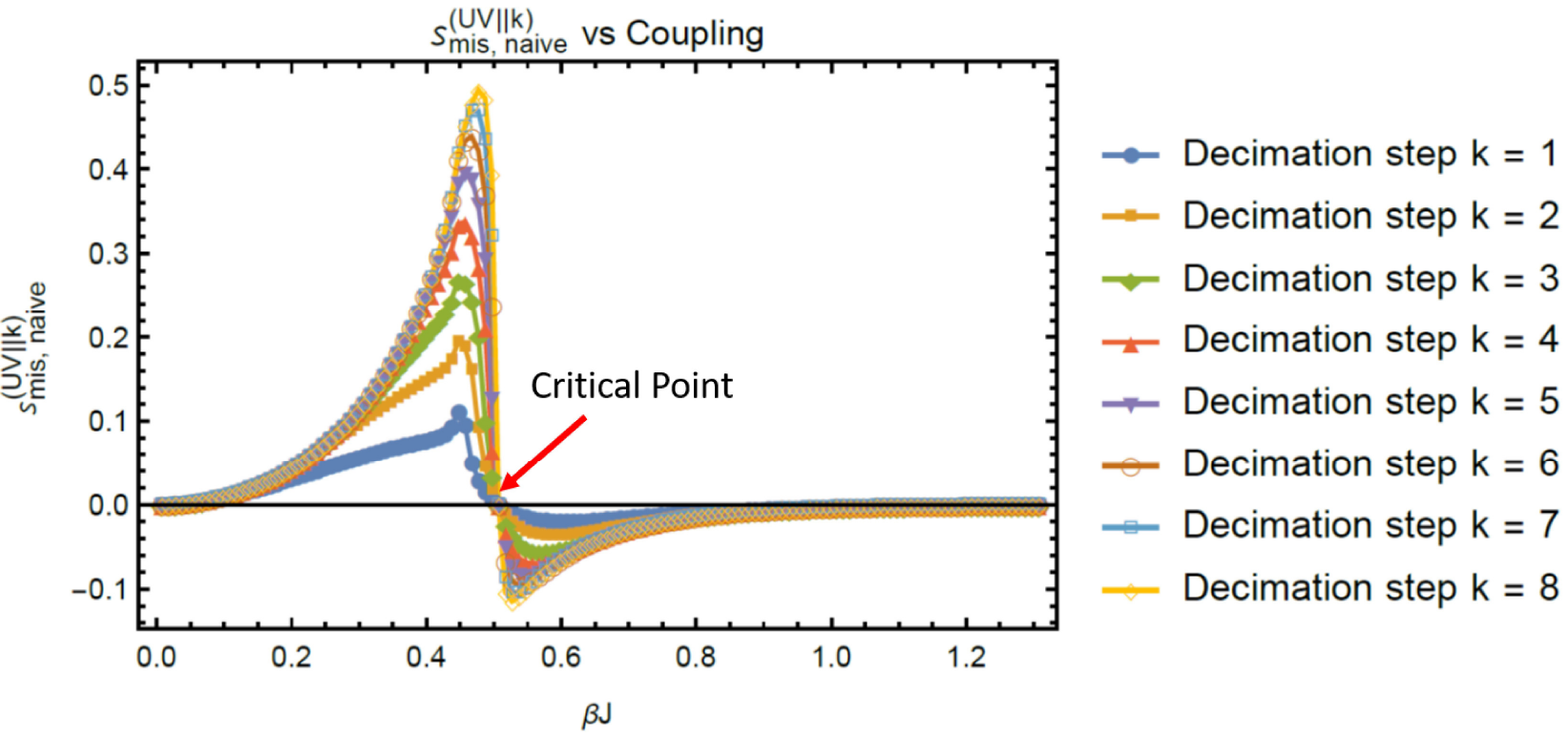}
\caption{The naive (in the sense that higher order interactions are dropped) approximation to the misanthropic entropy density $s_{\mathrm{mis,naive}}^{(\mathrm{UV}||k)}$ for the 2D Ising model as a function of the UV parameter $\beta J$ for various decimation steps. Only a few steps are shown because the curves merge to the exact same curve around decimation step 8, i.e., the approximation stops significantly changing after each decimation step after around eight decimation steps. The critical point in this approximation scheme is $\beta J_{\ast} \approx 0.50698$.}
\label{fig:kldens2Dvtemp1}
\end{figure}
\begin{figure}[t!]
\centering
\includegraphics[trim={0 4cm 0 4cm},scale=0.5]{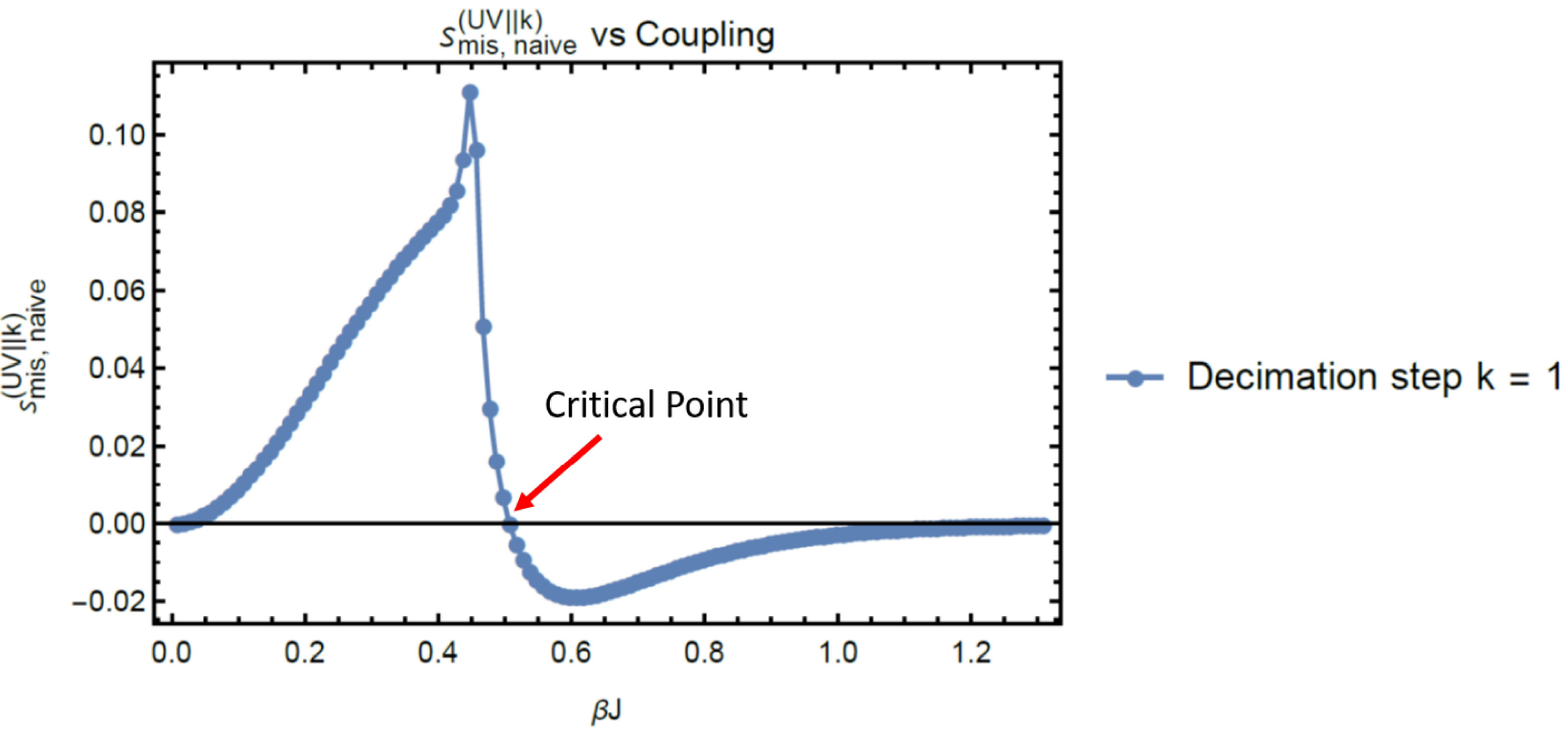}
\caption{The naive (in the sense that higher order interactions are dropped) approximate misanthropic entropy density $s_{\mathrm{mis,naive}}^{(\mathrm{UV}||1)}$ for the 2D Ising model as a function of the UV parameter $\beta J$ after one decimation step. On general grounds, the misanthropic entropy must be non-negative. The apparent negative values occur even after one decimation step, indicating that higher order interaction terms must be included to accurately compute the average energy after decimation. The critical point in this approximation scheme is $\beta J_{\ast} \approx 0.50698$.}
\label{fig:kldens2Dvtemp2}
\end{figure}

Taking $p$ to be the original distribution for the 2D Ising Model with $N$ sites and coupling $J$, with $q$ to be the joint distribution from multiplying the distributions of the decimated theory of $k$ steps with coupling $J_k$, we can write down a ``naive'' expression for the misanthropic entropy after $k$ decimation steps by dropping all quartic and higher order terms:
\begin{equation}
\begin{aligned}
\label{eq:2DDec3}
\mathcal{S}_{\mathrm{mis,naive}}^{(\mathrm{UV}||k)} = &2^k\log Z(N/2^k, J_k) - \log Z(N, J)\\
&+ 2^kE(N/2^k, J_k) - E(N, J),
\end{aligned}
\end{equation}
where $Z$ and $E$ are the 2D Ising model
partition function and average energy as found for example in many textbooks (see e.g. \cite{Pathria:1996hda}):
 \begin{equation}
\begin{aligned}
\label{eq:2DDec1}
\log Z &= N\log\left(\sqrt{2}\cosh(2\beta J)\right) + \frac{N}{\pi}\int_{0}^{\pi/2} \! \mathrm{d}\phi \, \, \log\left(1 + \sqrt{1 - \kappa^2\sin^2\phi} \right),\\
E \equiv E_{0} &= -\frac{\partial\log Z}{\partial\beta} = -NJ\coth(2\beta J)\left(1 + \frac{2}{\pi}K(\kappa)(2\tanh^2(2\beta J) - 1)\right),
\end{aligned}
\end{equation}
where $K$ is the complete elliptic integral of the first kind and $\kappa$ is given by:
\begin{equation}
\kappa = \frac{2\sinh(2\beta J)}{\cosh^2(2\beta J)}.
\end{equation}

Running through numerical values of $\beta J$, we can calculate the relative entropy density (we fix the number of sites to be $N = 2^{19}2^{19}$). Our results are in Figures \ref{fig:kldens2Dvtemp1}-\ref{fig:kldens2Dvtemp2}. Passing from small to large values of the initial UV coupling constant, we observe that there is a large peak and then drop as we move close to the critical point. The large increase is to be expected because the correlation length is increasing, and so the mutual information between nearest neighbors is becoming more prominent. Additionally, we observe that right at the critical point, our approximation vanishes. One reason to expect this in a numerical evaluation is that right at the critical point, the free energies satisfy:
\begin{equation}
F_{b,w}(\beta J_{\ast}) = F_{b}(\beta J_{\ast}) + F_{w}(\beta J_{\ast}),
\end{equation}
and so after using the relation $\mathcal{S} = - \partial F / \partial \beta^{-1}$ the misanthropic entropy formally vanishes.
Caution is warranted here because the entropies themselves are divergent at the critical point.

Continuing on to larger values of the coupling, observe that for $s_{\mathrm{mis,naive}}$,
the naive approximation to the misanthropic entropy obtained by dropping contributions from quartic and higher order terms to the Hamiltonian
would appear to produce negative values. This occurs even after one decimation step (Figure \ref{fig:kldens2Dvtemp2}), where we only have a quartic term to drop. The reason for the negative values comes from approximating the expectation value for the decimated Hamiltonian, i.e. by dropping the contribution from these quartic interactions to $E = \langle H \rangle$. This feeds into an approximation of the relative entropy rather than just an approximation of the distribution.

Note: Calculating the KL divergence numerically can sometimes result in a negative value. In our case, the negative values are not a result of the distributions being improperly normalized. The partition function and average energies are both calculated from normalized distributions. One of the distributions is a normalized distribution for a 2D Ising model with $N$ sites and coupling $J$. The other distribution is a 2D Ising model with $N/2^k$ sites and coupling constant $J_k$. To form the distributions in general, the steps we follow are to 1) decimate the Hamiltonian, 2) drop the quartic and higher interactions, 3) define the partition function in terms of the decimated Hamiltonian, i.e., the partition function $Z_k$ at some decimation step $k$ is computed by summing $\exp(-\beta H_k)$ (where $H_k$ is the Hamiltonian after dropping quartic and higher interactions) over all spin configurations. Our distributions are therefore automatically normalized, since the partition function that we use is defined in terms of the decimated Hamiltonian that drops the quartic and higher terms. Furthermore, since $H_k$ is also a 2D Ising model (having dropped higher terms), we can see that step 3) produces a partition function that is given by the approximation in equation \eqref{eq:2DDec1}, which can also be seen by inspecting the resulting partition function.

Indeed, on general information theoretic grounds we know that the full misanthropic entropy must be non-negative. Indeed, because we know that the only irrelevant operator is quartic at one decimation step, we also have some idea of the size of its contribution. Note also that other irrelevant operators can in principle make a significant contribution for greater decimation steps and the magnitude of the quartic coupling also changes with decimation step. Turning the discussion around, a negative value for the naive misanthropic entropy provides a way to \textit{detect} higher order interaction terms in the decimated theory!

\begin{figure}[t!]
\centering
\includegraphics[trim={0 4cm 0 4cm},scale=0.5]{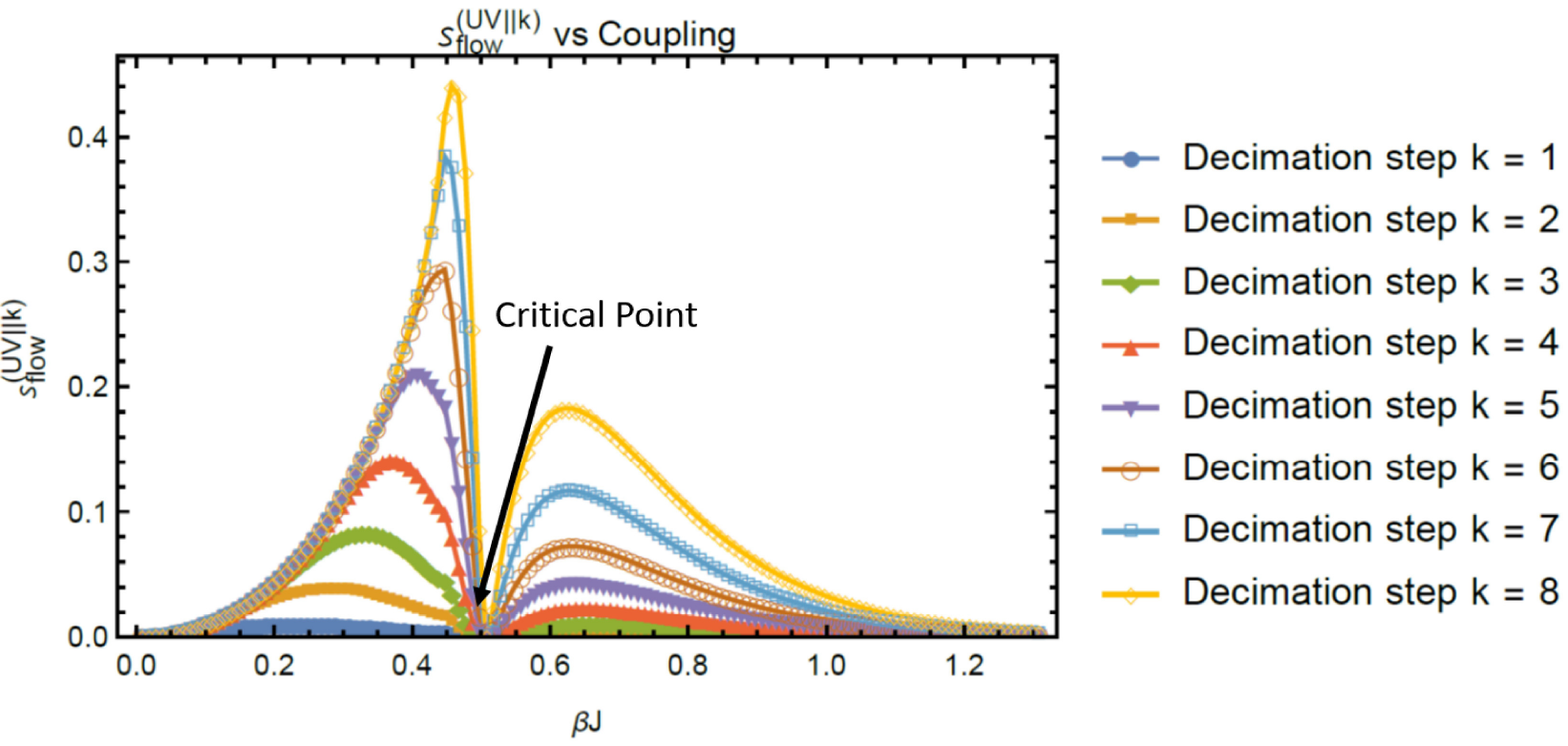}
\caption{The density $s^{(\mathrm{UV}||k)}_{\mathrm{flow}}$ for the 2D Ising model as a function of the UV parameter $\beta J$ for various decimation steps. Only a few steps are shown so as to easily compare the results with the results of $s_{\mathrm{mis}}$ in Figure \ref{fig:kldens2Dvtemp1}. Unlike in Figure \ref{fig:kldens2Dvtemp1}, the curves for $s^{(\mathrm{UV}||k)}_{\mathrm{flow}}$ continue to change with decimation, even all the way to the 19th decimation step (not shown here), with the humps above the critical value eventually becoming much larger than the humps before the critical value. The critical point in this approximation scheme is $\beta J_{\ast} \approx 0.50698$.}
\label{fig:kldens2DvtempRescaled1}
\end{figure}

\begin{figure}[t!]
\centering
\includegraphics[trim={0 4cm 0 4cm},scale=0.5]{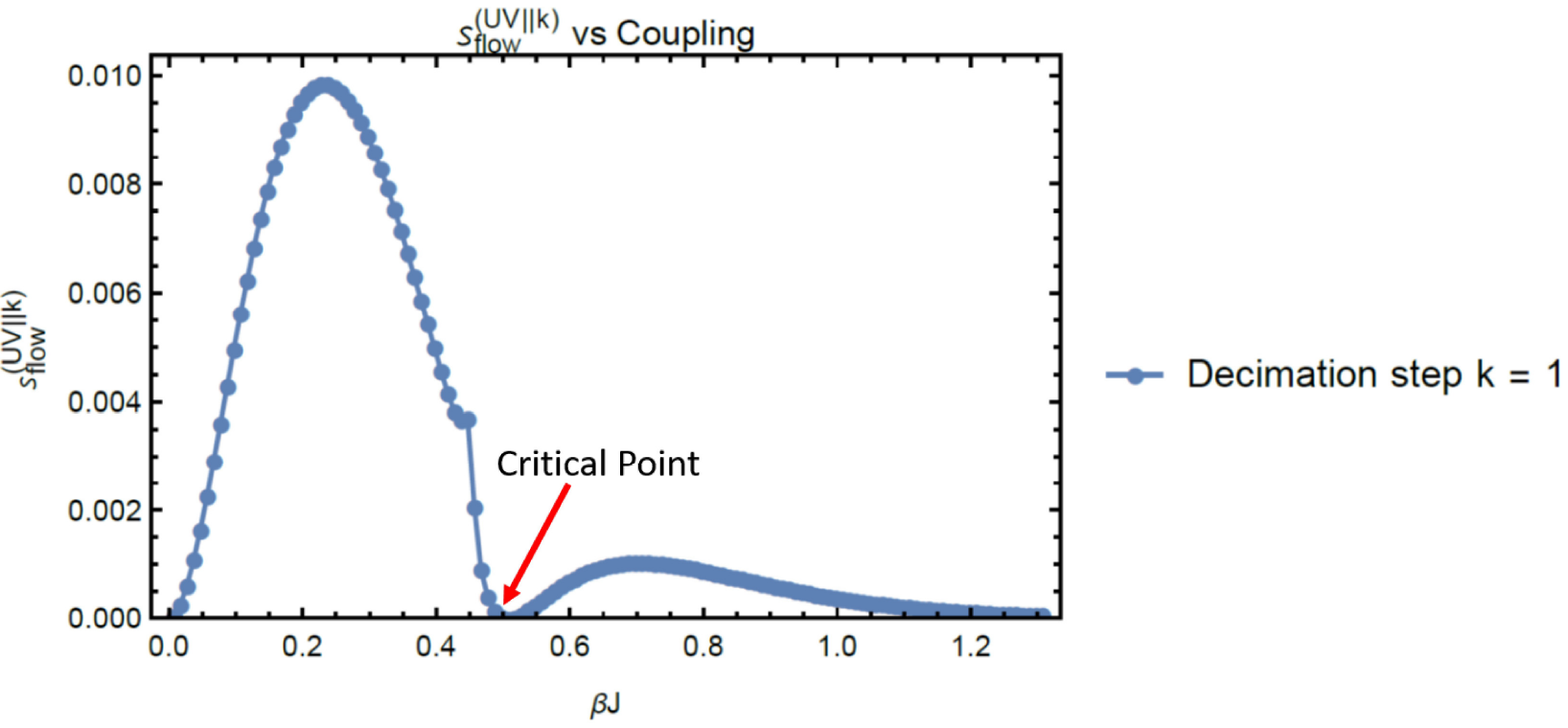}
\caption{The density $s^{(\mathrm{UV}||k)}_{\mathrm{flow}}$ for the 2D Ising model as a function of the UV parameter $\beta J$ after one decimation step. Compare with the result for $s_{\mathrm{mis}}$ in Figure \ref{fig:kldens2Dvtemp2}. The critical point in this approximation scheme is $\beta J_{\ast} \approx 0.50698$.}
\label{fig:kldens2DvtempRescaled2}
\end{figure}

As confirmation that the negative result comes from dropping irrelevant operators when approximating the expectation value of the Hamiltonian, consider that we could have also computed $s^{(\mathrm{UV}||k)}_{\mathrm{flow}}$, i.e., using the decimation plus rescaling procedure: comparing the original theory to the decimated theory on a rescaled lattice. For decimation plus rescaling, the rescaled Hamiltonian is (rescaling performed after dropping the quartic and higher interactions and combining the nearest and next-nearest neighbor interactions as usual),
\begin{equation}
\begin{aligned}
\label{eq:2DDecR1}
H_k  & \equiv -J_k\sum_{n.n.}\sigma_i\sigma_j,
\end{aligned}
\end{equation}
where the nearest neighbor sum goes over $N$ sites. That is, it is the exact same Hamiltonian as the $N$-site 2D Ising model with a different coupling constant. The quantity $\mathcal{S}_{\mathrm{flow}}^{(\mathrm{UV}||k)}$ becomes (again, we are dropping quartic and higher interactions),
\begin{equation}
\begin{aligned}
\label{eq:2DDecR2}
\mathcal{S}_{\mathrm{flow}}^{(\mathrm{UV}||k)} &= \log Z_k - \log Z + \left(-\langle H\rangle_{\mathrm{UV}} + \langle H_k\rangle_{\mathrm{UV}}\right),\\
\end{aligned}
\end{equation}
where we have,
\begin{equation}
\begin{aligned}
\label{eq:2DDecR3}
\langle H_k\rangle_{\mathrm{UV}} &= \frac{J_k}{J}\langle H\rangle_{\mathrm{UV}}.
\end{aligned}
\end{equation}
This follows because we can just pull the coupling constant $J$ out of the integral and evaluate the correlation function and then substitute in the $J_k$ coupling constant from the decimation procedure. This is just the same as multiplying the expectation value of the original Hamiltonian by $J_k/J$. Notice also in \eqref{eq:2DDecR2} that we have no factors of $2^k$--unlike with computing $s_{\mathrm{mis}}$ because there is no need to multiply copies of the decimated distribution together: the rescaled Hamiltonian has the same spin support as the original Hamiltonian. The factors of $2^k$ found in computing $s_{\mathrm{mis}}$ do not make a difference anyway: $\log Z_k$ and $E_k$ are directly proportional to the number of sites, so the $2^k$ from multiple copies of the decimated distribution cancels with the $N/2^k$ in front of these terms.

See Figures \ref{fig:kldens2DvtempRescaled1}-\ref{fig:kldens2DvtempRescaled2} for results and compare with the earlier corresponding Figures \ref{fig:kldens2Dvtemp1}-\ref{fig:kldens2Dvtemp2}. Observe that our approximation of $s_{\mathrm{flow}}$ is non-negative everywhere. This is  because our expression only approximates the distribution, not the entropy itself (which is different from what happens for $s_{\mathrm{mis}}$). Note also that as we vary the initial value of the coupling constant and tune it to the critical point, the value of $s_{\mathrm{flow}}$ vanishes. Away from this value, we observe a peak which grows more pronounced as we increase the number of decimation steps. This is to be expected because the fixed point of the Ising model is unstable.

As a final comment, one may wonder why the misanthropic entropy is not symmetric about the critical point.\footnote{We thank the journal referee for a prompting question on this point.} The Kramers-Wannier self-duality of the 2D square Ising model relates the high temperature regime to the low temperature regime, so why are the higher order operators behaving differently at high temperature vs low temperature? The answer is: the Kramers-Wannier duality does not merely exchange low and high temperature, but it also exchanges order and disorder operators. Our Hamiltonian consists of local spin operators and thereby favors the order operators. Another comment here is that the Krammers-Wannier self-duality exchanges the partition function up to a coupling-dependent factor; to see the duality on a plot, this factor must be taken into account when exchanging low and high temperatures. We leave the development of a suitable duality covariant entropy measure for future work.

\section{Conclusions}
\label{section:conclusion}

There is an intuitive sense in which information is lost under renormalization group flow.
In this note we recast this as the specification of a noisy communication channel,
with transmission of UV degrees of freedom to the IR via RG flow. We studied two entropic quantities
to track the flow of information, introducing a ``flow entropy'' $\mathcal{S}_{\mathrm{flow}}$
associated with treating RG flows as a trajectory in the space of couplings, and a ``misanthropic entropy''
$\mathcal{S}_{\mathrm{mis}}$ as associated with tracking the mutual information of nearest neighbors which are
about to be decimated. Infinitesimally, these two notions agree, but they track different properties in
long RG flows. We also explained how to define similar quantities for quantum statistical ensembles.
Using the 2D Ising model as an illustrative example, we showed that both quantities detect the onset of the critical
point, and moreover, the misanthropic entropy can, in naive approximations where higher order terms are dropped,
seem to be negative. This latter feature can actually be
used as a diagnostic to improve a candidate decimation procedure. In the remainder of this section we discuss some future
potential avenues for investigation.

Previous work in \cite{Balasubramanian:2014bfa} has shown that with a 2D CFT, a notion of relative entropy related to the misanthropic and ``flow'' entropies reduces, in the infinitesimal limit, to the Zamolodchikov metric \cite{Zamolodchikov:1986gt}.
It would be interesting to see how the Zamolodchikov metric relates to the misanthropic and flow entropies and whether and how the c-theorem relates to these quantities.

It would be interesting to consider more general choices of decimation procedure, and to track the sense in which
one choice or another leads to a minimal distortion rate. A related question is to consider varying the possible UV theories,
so as to optimize the transmission of information from the UV to the IR. This would seem to be an important question in determining
the extent to which a person confined to the IR can ever probe the UV.

Though we have primarily focused on the case of statistical field theories, many of the structures present here
generalize to the quantum setting. For example, it would likely be illuminating to consider in detail the case of the
quantum Ising spin chain in the presence of a transverse magnetic field, which is known to enjoy a quantum critical point (for a review, see e.g. \cite{sachdev_2011}).

Lastly, the specification of renormalization as a communication channel is of course reminiscent of various information theoretic approaches to
aspects of the AdS/CFT correspondence, which at this point has grown to a vast literature. In that context, motion in the radial direction from the boundary into the interior roughly amounts to moving from the UV to the IR. This fits with the perspective of starting with a message encoded in the degrees of freedom of the CFT, and tracking what happens if we attempt to package this in terms of a coarse graining operation, much as in \cite{Swingle:2009bg}, and it would be interesting to spell this out in more detail.

As a particularly important case, observe that we have implicitly been considering an approximation to the thermal density matrix $\rho \sim \exp(- \beta \widehat{H})$. For a large $N$ CFT, the resulting thermofield double state is expected to be dual to an eternal black hole in anti-de Sitter space \cite{Maldacena:2001kr}. It would be interesting to study the impact of Weyl rescalings on such density matrices and the corresponding holographic interpretation of misanthropic entropy.


\section*{Acknowledgements}

We thank J. Drut, V. Oganesyan, J. Porter, D. Radi\v{c}evi\'{c} and D. Schwab for helpful discussions.
JJH thanks the 2021 Simons summer workshop at the Simons Center for Geometry and Physics
for kind hospitality during the completion of this work. The work of JJH is
supported by the DOE (HEP) Award DE-SC0013528.

\newpage


\bibliographystyle{utphys}
\bibliography{UVIR}

\end{document}